\documentclass{elsart3p}
\usepackage{times,mathptmx}
\usepackage[numbers,sort&compress]{natbib}
\usepackage{amssymb}
\usepackage[final]{graphicx}
\usepackage{units}
\usepackage{amsmath}
\def\lsi{\raise0.3ex\hbox{$<$\kern-0.75em\raise-1.1ex\hbox{$\sim$}}}
\def\gsi{\raise0.3ex\hbox{$>$\kern-0.75em\raise-1.1ex\hbox{$\sim$}}}

\newcommand{\gsim}{\mathop{\gsi}}
\def\be{\begin{equation}}
\def\ee{\end{equation}}
\def\ba{\begin{eqnarray}}
\def\ea{\end{eqnarray}}

\begin{document}

%%%%%%%%%%%%%%%%%%%%%%%%%%%%%%%%%%%%%%%%%%%%%%%%%%%%%%%%%%%%%%%%%%%%%%

\begin{frontmatter}
\title{Quantum scale invariance, cosmological constant and hierarchy
problem}

\author{Mikhail Shaposhnikov}
\ead{Mikhail.Shaposhnikov@epfl.ch} and
\author{Daniel Zenh\"ausern}
\ead{Daniel.Zenhaeusern@epfl.ch}

\address{
  Institut de Th\'eorie des Ph\'enom\`enes Physiques,
  \'Ecole Polytechnique F\'ed\'erale de Lausanne,
  CH-1015 Lausanne, Switzerland}

\date{14 November 2008}

\begin{abstract}

We construct a class of theories which are scale-invariant on quantum
level in all orders of perturbation theory. In a subclass of these
models scale invariance is spontaneously broken, leading to
the existence of a massless dilaton. The applications of these
results to the problem of stability of the electroweak scale against
quantum corrections, to the cosmological constant problem and to
dark energy are discussed.

\end{abstract}


\end{frontmatter}

%%%%%%%%%%%%%%%%%%%%%%%%%%%%%%%%%%%%%%%%%%%%%%%%%%%%%%%%%%%%%%%%%%%%%%

\section{Introduction}
\label{sec:intro}
If in any theory all dimensionfull parameters (generically denoted by
$M$), including masses of elementary particles, Newton's gravitational
constant, $\Lambda_{QCD}$ and alike are rescaled by the same amount $M
\to M \sigma$, this cannot be measured by any observation. Indeed,
this change, supplemented by a dilatation of space-time coordinates
$x^\mu \to \sigma x^\mu$ and an appropriate redefinition of the fields
does not change the complete quantum effective action of the theory.
However, the  symmetry transformations
in quantum field theory only act on fields and not on parameters of
the Lagrangian. The realization of scale invariance happens to
be a non-trivial problem. A classical field
theory which does not contain any dimensionfull parameters is
invariant under the substitution 
\be 
\Phi(x) \to \sigma^n\Phi(\sigma x)~,
\ee 
where $n$ is the canonical mass dimension of the field $\Phi$. This
dilatational symmetry turns out to be anomalous on quantum level for
all
realistic renormalizable quantum field theories (for a review see
\cite{coleman}). The divergence of the dilatation current $J_\mu$ is
non-zero and is proportional to the $\beta$-functions of the
couplings. For example, in pure gluodynamics, scale-invariant on the
classical level, one has
\be
\partial_\mu J^\mu \propto \beta(g) G_{\alpha\beta}^a
G^{\alpha\beta~a}~,
\ee
where $G_{\alpha\beta}^a$ is the non-Abelian gauge field strength.

At the same time, it is very tempting to have a theory which is
scale-invariant (SI) on the quantum level, as this would solve a
number of
puzzles in high energy physics. Most notably, these problems include
two tremendous fine-tunings, facing the Standard Model (SM). The first
one is related to the stability of the Higgs mass against radiative
corrections and the second one to the cosmological constant problem.
If the full quantum theory, including gravity, is indeed
scale-invariant, and SI is broken spontaneously, the Higgs mass is
protected from radiative corrections by an exact dilatational
symmetry.

Moreover, as we have shown in  \cite{Shaposhnikov:2008xb}, the {\em
classical} theory with SI broken spontaneously and given by the action
(we omit from the Lagrangian of \cite{Shaposhnikov:2008xb} all degrees
of freedom which are irrelevant for the present discussion and keep
only the gravity part, the Higgs field $h$ and the dilaton $\chi$):
\be
{\cal L}_{tot} = {\cal L}_G + {\cal L}~,
\label{ltot}
\ee
where
\begin{eqnarray}
\label{LG} 
&&{\cal L}_G=  -\left(\xi_\chi \chi^2 + \xi_h h^2\right)\frac{R}{2}~,\\
&&{\cal L}=\frac{1}{2}\left[(\partial_\mu\chi)^2 +(\partial_\mu h)^2
\right]-\lambda\left(h^2-\zeta^2\chi^2\right)^2~,
 \label{L}
\end{eqnarray}
not only has zero cosmological constant but also gives a source for
dynamical dark energy, provided that gravity is unimodular, i.e. the
determinant of the metric is fixed to be $-1$. (Here $R$ is the scalar
curvature and $\xi_\chi,~\xi_h,~\lambda$ and $\zeta$ are dimensionless
coupling constants.) In this theory all mass parameters (on the tree
level) come from one and the same source -- the vacuum expectation
value of the dilaton field $\langle\chi\rangle = \chi_0$, which is
exactly massless. In addition, the primordial inflation is a natural
consequence of (\ref{ltot}), with a Higgs field playing the role of
the inflaton \cite{Bezrukov:2007ep}.

It looks like all these findings are ruined by quantum corrections.
The aim of this Letter is to show that this is not the case. We will
construct a class of effective field theories, which obey the
following properties:\\
(i) Scale invariance is preserved on quantum level in all orders of
perturbation theory.\\
(ii) Scale invariance is broken spontaneously, leading to a massless
dilaton.\\
(iii) The effective running of coupling constants is automatically
reproduced at low energies.\\
In other words, the benefits of classical SI theories (no corrections
to the Higgs mass, zero cosmological constant, presence of dark energy
and primordial inflation) can all be present on the quantum level. At
the same time, the standard results of quantum field theory, such as
the running of coupling constants, remain in place. Whether the
theories we construct are renormalizable\footnote{The precise sense of
this word in the present context will be specified later.} and unitary
is not known to us (though we will formulate some conjectures on this
point). However, the renormalizability is not essential for the
validity of the results.

The Letter is organized as follows. In Section \ref{sec:example} we
explain our main idea with the use of a simple model of two scalar
fields. In Section \ref{sec:scale} we describe its generalization to
an arbitrary case. In Section \ref{sec:conjecture} we discuss the
inclusion of gravity and present our conclusions in Section \ref
{sec:concl}.

%%%%%%%%%%%%%%%%%%%%%%%%%%%%%%%%%%%%%%%%%%%%%%%%%%%%%%%%%%%%%%%%%%%%%%
\section{Scalar field example}
\label{sec:example}
We will explain our idea using the example of a simple system
containing two scalar fields described in classical theory by the
Lagrangian (\ref{L}) without gravity. The construction is essentially
perturbative and based on the dimensional regularization of 't Hooft
and Veltman \cite{'tHooft:1972fi} (for a discussion of the hierarchy
problem within this scheme see, e.g. \cite{Shaposhnikov:2007nj}).

At the classical level the theory (\ref{L}) is scale-invariant. In
fact, the requirement of dilatational symmetry  does not forbid the
presence of an additional term $\beta \chi^4$ in \eqref{L}. If $\beta
<0$, the theory does not have a stable ground state, for $\beta>0$ the
ground state is unique and corresponds to $h=\chi=0$. At the classical
level one would conclude that the theory contains two scalar massless
excitations for the ground state respecting scale invariance. For
the case  $\beta=0$ the potential contains two flat directions $h =
\pm \zeta \chi$ and the vacuum is degenerate. If $\chi =\chi_0 \neq
0$, the dilatational invariance is spontaneously broken. Then the
theory
contains a massive Higgs boson, $m_H^2(\chi_0) = 2\lambda
\zeta^2(1+\zeta^2)\chi_0^2$ and a massless dilaton. So, the only
choice for $\beta$, interesting for phenomenology, is $\beta=0$,
otherwise the vacuum does not exist or the theory does not contain any
massive particles\footnote{More discussion of the $\beta>0$ case will
be 
given at the end of this section.}. In what follows we will also
assume that $\zeta \lll 1$, which is true for phenomenological
applications: ${\chi_0 \sim M_P= 2.44\times 10^{18}}$ GeV is related
to
the Planck scale, and ${h_0 =\zeta \chi_0 \sim M_W \sim 100}$ GeV to
the
electroweak scale. However, the smallness of $\zeta$ is not essential
for the theoretical construction. 

It is well known what happens in this theory if the {\em standard}
renormalization procedure is applied. In  $d$-dimensional space-time
(we use the convention ${d=4-2\epsilon}$) the mass dimension of the
scalar fields is $1-\epsilon$, and that of the coupling constant
$\lambda$ is $2\epsilon$. Introducing a (finite) dimensionless
coupling $\lambda_R$, one can write
\be
\lambda = \mu^{2\epsilon}\left[\lambda_R + \sum_{n=1}^\infty
\frac{a_n}{\epsilon^n}\right]~,
\label{hooft}
\ee
where $\mu$ is a dimensionfull parameter and the Laurent series in
$\epsilon$  corresponds to counter-terms. The parameters $a_n$ are to
be fixed by the requirement that renormalized Green's functions are
finite in every order of perturbation theory. Similar replacements are
to be done with other parameters of the theory, and the factors
$Z_\chi,~Z_h$, related to the renormalization of fields must be
introduced (they do not appear at one-loop level in our scalar
theory). Then, in the $\overline{MS}$ subtraction scheme, the one-loop
effective potential along the flat direction has the form
\be
V_1(\chi)=
\frac{m_H^4(\chi)}{64\pi^2}\left[\log\frac{m_H^2(\chi)}{
\mu^2 }
-\frac{3}{2}\right]~,
\ee
spoiling its degeneracy, and leading thus to explicit breaking of the
dilatational symmetry. The vacuum expectation value of the field
$\chi$ can be fixed by renormalization conditions
\cite{Coleman:1973jx}. The dilaton acquires a nonzero mass. It is the
mismatch in mass dimensions of bare ($\lambda$) and renormalized
couplings ($\lambda_R$) which leads to the dilatational anomaly and
thus to
explicit breaking of scale invariance (see  \cite{Meissner:2006zh}
for a recent discussion).

Let us now use another prescription, which we will call the ''SI
prescription''\footnote{A similar procedure was suggested in
\cite{Englert:1976ep} in connection with the conformal anomaly. We
thank Thomas Hertog who pointed out this reference to us after our
work has been submitted to hep-th.}. Replace $\mu^{2\epsilon}$ in
(\ref{hooft}) and in all other similar relations by (different, in
general) combinations of fields $\chi$ and $h$, which have the correct
mass dimension:
\be
\mu^{2\epsilon} \to 
\chi^{\frac{2\epsilon}{1-\epsilon}}F_\epsilon(x)~,
\label{new}
\ee
where $x = h/\chi$ and $F_\epsilon(x)$ is a function depending on the
parameter $\epsilon$ with the property $F_0(x)=1$. In principle, one
can use different functions $F_\epsilon(x)$ for the various couplings.
The resulting field theory, by construction, is scale-invariant for
any number of space-time dimensions $d$. This means, that if for
instance the $\overline{MS}$ subtraction scheme is used for
calculations, the renormalized theory is also scale-invariant in any
order of perturbation theory. 

The requirement of scale invariance itself does not fix the details of
the prescription. However, the form of the couplings of the scalar
fields $\chi$ and $h$ to gravity as in Eq. (\ref{LG}) indicates that
the combination  
\be
\xi_\chi \chi^2 + \xi_h h^2 \equiv
\omega^2
\ee
plays a special role, being the effective Planck constant. Therefore,
we arrive to a simple  ``GR-SI prescription'', in which
\be
\mu^{2\epsilon} \to \left[\omega^2\right]^{\frac{\epsilon}{1-\epsilon}}~,
\label{GR}
\ee
corresponding to the choice of the
function $F_\epsilon(x)=(\xi_\chi+\xi_h
x^2)^{\frac{\epsilon}{1-\epsilon}}$.
We will apply the GR-SI prescription to the one-loop analysis of our
scalar theory
below. In the appendix we will consider a modified variant of the 
procedure.

The SI construction is entirely perturbative and can in fact be used
{\em only} if SI is spontaneously broken. In other words, in order to
use the GR-SI prescription the ground state has to be
$(h_0,\chi_0)\neq (0,0)$, because otherwise it is impossible to
perform an expansion of (\ref{GR}). Indeed, consider the exact
effective potential $V_{eff}(h,\chi)$ of our theory, constructed using
the prescription (\ref{new}) or (\ref{GR}) in the limit $\epsilon \to
0$. Because of exact SI, it can be written
as 
\be
V_{eff}(h,\chi) = \chi^4 V_{\chi}(x)=h^4 V_{h}(x)~.
\label{form}
\ee
For the ground state to exist, we must have $V_{\chi}(x)~\geq~0$ (or,
what is the same, $V_{h}(x)~\geq~0$) for all $x$. For the minimum of
$V_{eff}(h,\chi)$ to lie in the region where $\chi\neq 0$ (or $h\neq
0$), we must have $V_\chi(x_0)=0$ ($V_h(x_0)=0$), where $x_0$ is a
solution of $V_\chi'(x_0)=0$ ($V_h'(x_0)=0$) and prime denotes the
derivative with respect to $x$. If these conditions are satisfied, the
theory has an infinite set of ground states corresponding to the
spontaneous breakdown of dilatational invariance. The dilaton is
massless in all orders of perturbation theory. In this case one can
develop the perturbation theory around the vacuum state corresponding
to $\chi_0~\neq~0$, $h_0=x_0\chi_0$ with arbitrary $\chi_0$ (or
$h_0~\neq~0$, $\chi_0=h/x_0$ with arbitrary $h_0$).

To summarize: the use of prescriptions (\ref{new}) or (\ref{GR})
supplemented by the requirement $V_{\chi,h}(x_0)=0$ leads to a new
class of theories exhibiting spontaneously broken scale invariance,
which is exact on quantum level. These theories can be called
renormalizable if the introduction of a finite number of counter-terms
is sufficient to remove all divergences and guarantee the existence of
a flat direction in the potential. The check whether this is indeed
the case goes beyond the scope of the present Letter. In principle, we
cannot exclude the possibility that, in order to remove all
divergences, a new type of counter-terms containing non-polynomial
interactions (such as  $h^6/\chi^2$) is required. But, even if this is
the case, scale invariance is maintained in all orders of perturbation
theory and can be spontaneously broken. Another potential issue is
unitarity. We do not know whether higher derivative terms in the
effective action, dangerous from this point of view, would require the
introduction of corresponding counter-terms. However, the {\em
functional} arbitrariness in the choice of $F_\epsilon(x)$ for
potential and
kinetic terms may give enough freedom to remove the unwanted
contributions. 

The theories we construct are quite different from ordinary
renormalizable theories. Their physics is determined not only by the
values of ``classical'' coupling constants ($\lambda$ and $\zeta$ in
our case), but also by ``hidden'' parameters contained in the
functions $F_{\epsilon}(x)$. Still, as we will see shortly, for the
SI-GR prescription, in the limit $\zeta \lll 1$ and for small energies
$E \ll \chi_0$, only ``classical'' parameters matter. Moreover, they
automatically acquire the necessary renormalization group running.

To this end, we carry out a one-loop analysis of the theory (\ref{L})
with the GR-SI prescription. We write the $d$-dimensional
generalization of the classical potential as\footnote{If we define the
parameters 
$\alpha\equiv\sqrt{\lambda}$
 and
$\beta\equiv\sqrt{\lambda}\zeta^2$,
 the classical potential takes the form 
$U=\frac{1}{4}\left(\alpha h^2-\beta \chi^2\right)^2$. 
 In this notation the GR-SI prescription corresponds to the substitutions
${\alpha\to\left[\omega^2\right]^{\frac{\epsilon}{2(1-\epsilon)}}
\alpha_R}$
 and
${\beta\to\left[\omega^2\right]^{\frac{\epsilon}{2(1-\epsilon)}}
\beta_R}$.}
\be
U=\frac{\lambda_R}{4}\left[\omega^2\right]^{\frac{\epsilon}{1-\epsilon}}
\left[h^2-\zeta_R^2 \chi^2\right]^2\;,
\label{UUU}
\ee
and introduce the counter-terms
\ba
\label{cc}
U_{cc}=
\left[\omega^2\right]^{\frac{\epsilon}{1-\epsilon}}
\Bigg[&&A h^2 \chi^2 \left(\frac{1}{\bar\epsilon}+a\right)+\\ 
&&B \chi^4     \left(\frac{1}{\bar\epsilon}+b\right) + 
C h^4        \left(\frac{1}{\bar\epsilon}+c\right)\Bigg] ~,
\nonumber
\ea
where $\frac{1}{\bar \epsilon}=\frac{1}{\epsilon}-\gamma+\log(4\pi)$,
$\gamma$ is the Euler constant and $a,~b,~c,~A,~B$, and $C$ are
arbitrary for the moment.  We do not introduce any modification of the
kinetic terms since no wave function renormalization is expected at
the one loop level.

It is straightforward to find the one-loop effective potential for
this theory. The counter-terms removing the divergences coincide with
those of the standard prescription and are given by:
\ba
\nonumber
&&A~\to -\lambda_R^2 \zeta_R^2\frac{3 \zeta_R^4-4\zeta_R^2+3}{32 \pi^2}~,\\
&&B~\to~ \lambda_R^2 \zeta_R^4\frac{9\zeta_R^4+1}{64\pi^2}~,~~~
C~\to~ \lambda_R^2 \frac{\zeta_R^4+9}{64\pi^2}~.
\label{ABC}
\ea
The potential itself has a generic form $U_1=\chi^4 W_1(x)$ and is
given by a rather lengthy expression (we do not present it here, since
it is not very illuminating), which also depends on the ``hidden''
parameters. For a generic choice of $a,~b,$ and $c$ the classical flat
direction $x_0=\zeta_R$ is lifted by quantum effects. However, the
requirement $W_1(\zeta_R)=W'_1(\zeta_R)=0$ allows to fix  two of these
parameters  in a way such that the one-loop potential has exactly the
same flat direction. For $\zeta_R \lll 1$ this requirement leads
to\footnote{The truncation only serves to shorten the expressions.
There is no difficulty in finding the exact relations.}
\ba
\nonumber
&&b=3a +2\log\left(\frac{2\lambda_R\zeta_R^2}{\xi_\chi}\right)
+{\cal O}\left(\zeta_R^2\right)~,\\
&&c=\frac{1}{3}\left[a+2-2\log
\left(\frac{2\lambda_R\zeta_R^2}{\xi_\chi}\right)
\right]
+{\cal O}\left(\zeta_R^2\right)~.
\ea
The function $W_1(x)$ is positive near the flat direction, provided 
$a+2+2\log\left(\frac{2\lambda_R\zeta_R^2}{\xi_\chi}\right)>0$.

It is interesting to look at the one-loop effective potential as a
function of $h$ for $\chi=\chi_0$, $h \sim \zeta_R \chi_0 \equiv v$ and
$\zeta\lll 1$,
i.e. $h_0\lll \chi_0$.  One finds
\ba
\label{pp}
U_1 = &&\frac{m^4(h)}{64\pi^2}\left[\log\frac{m^2(h)}{v^2}
+{\cal O}\left(\zeta_R^2\right)\right]\\
+&&\frac{\lambda_R^2}{64\pi^2}
\left[C_0 v^4 + C_2 v^2 h^2 + C_4 h^4\right]
 + {\cal O}\left(\frac{h^6}{\chi^2}\right),
\nonumber
\ea
where $m^2(h)=\lambda_R (3h^2-v^2)$ and 
\ba
\nonumber
C_0&&=~~\frac{3}{2}\left[2a-1 + 
2\log\left(\frac{\zeta_R^2}{\xi_\chi}\right)+
\frac{4}{3}\log2\lambda_R+O(\zeta_R^2)\right]~,\\
\label{const}
C_2&&=-3\left[2a-3 + 
2\log\left(\frac{\zeta_R^2}{\xi_\chi}\right)+O(\zeta_R^2)\right]~,\\
C_4&&=~~\frac{3}{2}\left[2a-5 + 
2\log\left(\frac{\zeta_R^2}{\xi_\chi}\right)-
4\log2\lambda_R+O(\zeta_R^2)\right]~.
\nonumber 
\ea
The first term  in (\ref{pp}) is exactly the standard effective
potential for the theory (\ref{L}) with the dynamical field $\chi$
replaced by a constant $\chi_0$, while the rest is a quartic
polynomial of $h$ and  comes from our GR-SI prescription, leading to
redefinition of coupling constants, masses, and the vacuum energy.

One can see from \eqref{pp} that the quantum corrections to the Higgs
mass are proportional to $v^2\propto\zeta_R^2\chi^2$. This means that
they are small compared to the classical value. Moreover, the
potentially dangerous corrections of the type
$\lambda^n\chi_0^2$ to the Higgs mass {\em cannot} appear in higher
orders of perturbation theory. Indeed, for $\zeta=0$ the Higgs field
decouples from the dilaton at the classical level and the dilaton
field is described by a free theory. Therefore, if $\zeta=0$, the
value of the
(large) field $\chi$ can appear only through log's in the effective
potential, coming from the expansion of
$[\omega^2]^{{\epsilon}/{1-\epsilon}}$ in Eq. \ref{UUU}, or at
most as
$\zeta_R^2\chi^2$ if $\zeta\neq 0$. Hence, in this theory there is no
problem of instability of the Higgs mass against quantum corrections, 
appearing in the Standard Model.

Consider now the high energy (${\sqrt{s}\gg v}$ but
${\sqrt{s}\ll\chi_0}$)
behaviour of scattering amplitudes with the example of Higgs-Higgs
scattering (assuming, as usual, that ${\zeta_R\lll1}$). It is easy to
see
that in one-loop approximation one gets for the 4-point function
\be
\Gamma_4 = \lambda_R +
\frac{9\lambda_R^2}{64\pi^2}\left[\log\left(\frac{s}{\xi_\chi\chi_0^2}\right)
+{\rm const}\right]
+{\cal O}\left(\zeta_R^2\right)~.
\label{RG}
\ee
This implies that at $v\ll\sqrt{s}\ll\chi_0 $ the effective Higgs 
self-coupling  runs in a way prescribed by the ordinary
renormalization group. Not only the tree Higgs mass is determined by
the vev of the dilaton, but also all $\Lambda_{QCD}$-like parameters.
We expect that these results remain valid in higher orders of
perturbation theory.

Let us comment now on the case when the flat direction does not exist
at the quantum level (classically this corresponds to $\beta>0$). Then
the ground state of the theory is scale-invariant. Theories of this
type do not in general contain asymptotic
particle states (for a
review see, e.g. \cite{Aharony:1999ti}). If they do (this would
correspond to anomalous dimensions for the fields equal to zero), the
propagators will coincide with the free ones, leading to a theory with
a trivial S-matrix  \cite{Greenberg:1961mr,Bogolubov:1975}. In other
words, the requirement that the scale-invariant quantum field theory
can be used for the description of interacting particles, existing as
asymptotic states, singles out the class of theories with spontaneous
breaking of scale invariance. 

%%%%%%%%%%%%%%%%%%%%%%%%%%%%%%%%%%%%%%%%%%%%%%%%%%%%%%%%%%%%%%%%%%%%%%
\section{Scale-invariant quantum field theory: General formulation}
\label{sec:scale}
It is straightforward to generalize the construction presented above
to the case of theories containing fermions and gauge fields, such as
the Standard Model. The mass dimension of a fermionic field is
$\frac{3}{2}-\epsilon$, leading to the dimension of bare Yukawa
couplings $F_B$ equal to $\epsilon$. The mass dimension of the gauge
field can be fixed to $1$ for any number of space-time dimensions $d$,
leading to the dimensionality of the bare gauge coupling $g_B$  equal
to $\epsilon$. So, in the standard procedure one chooses $F_B \propto
\mu^\epsilon F_R,~g_B \propto \mu^\epsilon g_R$, where the index $R$
refers to renormalized couplings. For the SI or GR-SI prescription one
replaces $\mu^\epsilon$ by a combination of scalar fields of
appropriate dimension, as in (\ref{new}) or in (\ref{GR}). For the
perturbation theory to make sense, one has to choose counter-terms
in such a way that the full effective potential has a flat direction
corresponding to spontaneously broken dilatational invariance.

%%%%%%%%%%%%%%%%%%%%%%%%%%%%%%%%%%%%%%%%%%%%%%%%%%%%%%%%%%%%%%%%%%%%%%
\section{Inclusion of gravity}
\label{sec:conjecture}
The inclusion of scale-invariant gravity is carried out precisely
along the same lines. The metric tensor $g_{\mu\nu}$ is dimensionless
for any number of space-time dimensions and $R$ always has mass
dimension $2$. Therefore, the non-minimal couplings $\xi_\chi,~ \xi_h$
(see Eq. (\ref{LG})) are dimensionless and thus can only be multiplied
by functions $F_\epsilon(x)$ of the type defined in (\ref{new}). In
addition to (\ref{LG}), the gravitational action may contain the
operators $R^2,~R_{\mu\nu}R^{\mu\nu},~\Box R$ and
$R_{\mu\nu\rho\sigma}R^{\mu\nu\rho\sigma}$, multiplied by
$\chi^{\frac{-2\epsilon}{1-\epsilon}}F_\epsilon(x)$ (here
$R_{\mu\nu}$  and $R_{\mu\nu\rho\sigma}$ are the  Ricci and Riemann
curvature tensors). These operators are actually needed for
renormalization of field theory in curved space-time (for a review see
\cite{Buchbinder:1992rb}).

The presence of gravity is crucial for phenomenological applications.
Since Newton's constant is dynamically generated, the dilaton
decouples from the particles of the Standard model
\cite{vanderBij:1993hx,CervantesCota:1995tz,Wetterich:1987fk,
Wetterich:1987fm,Shaposhnikov:2008xb}, and thus satisfies all laboratory and
astrophysical constraints. As we found in \cite{Shaposhnikov:2008xb}, if gravity is
unimodular, the absence of a cosmological constant and the existence
of dynamical dark energy are automatic consequences of the theory. It
is interesting to note that the action of unimodular gravity is
polynomial with respect to the metric tensor. This leads us to the
conjecture that the SI unimodular gravity with matter fields may
happen to be a renormalizable theory in the sense described in
Section  \ref{sec:example}.

%%%%%%%%%%%%%%%%%%%%%%%%%%%%%%%%%%%%%%%%%%%%%%%%%%%%%%%%%%%%%%%%%%%%%%
\section{Conclusions}
\label{sec:concl}
In this Letter we constructed a class of theories, which are
scale-invariant on the quantum level. If dilatational symmetry is
spontaneously broken, all mass scales in these models are generated
simultaneously and originate from one and the same source. In these
theories the effective cutoff scale depends on the background dilaton
field, as was already proposed in \cite{Wetterich:1987fm}, which is
essential for inflation \cite{Bezrukov:2007ep} and dark energy
\cite{Shaposhnikov:2008xb}. The cosmological constant is absent and
the mass of the Higgs boson is protected from large radiative
corrections by the dilatational symmetry. Dynamical dark energy is a
remnant of initial conditions in unimodular gravity. 

There are still many questions to be understood. Here is a partial
list of them. Our construction is essentially perturbative. How to
make it work non-perturbatively\footnote{A proposal based on lattice
regularization has been discussed recently in
\cite{Shaposhnikov:2008ar}.}?  Though the stability of the electroweak
scale against quantum corrections is achieved, it is absolutely
unclear {\em why} the electroweak scale is so much smaller than the
Planck scale (or why $\zeta \lll 1$). It remains to be seen if this
new class of theories is renormalizable and unitary (note, though,
that renormalizability is not essential for the construction). At
large momentum transfer $p \gsim M_P$ the perturbation theory diverges
and thus is not applicable. What is the high energy limit of these
theories?

%%%%%%%%%%%%%%%%%%%%%%%%%%%%%%%%%%%%%%%%%%%%%%%%%%%%%%%%%%%%%%%%%%%%%%
{\bf Acknowledgements.}
This work was supported by the Swiss National Science Foundation.
We thank F. Bezrukov, K. Chetyrkin, S. Sibiryakov  and I. Tkachev for valuable comments.

\section*{Appendix}
For the GR-SI prescription considered in the Letter, physics well
below
the Planck scale associated with the dilaton vev $\chi_0$ was the same
as for the ordinary renormalizable scalar theory containing the Higgs
field $h$ only. This is not necessarily the case if the SI
prescription
given by Eq. (\ref{new}) is used. Indeed, consider now a distinct way
of continuing the scalar potential to $d$-dimensional
space-time:\footnote{In the notation with $\alpha\equiv\sqrt{\lambda}$
and
$\beta\equiv\sqrt{\lambda}\zeta^2$, the prescription used
here corresponds to the substitutions
$\alpha\to h^{\frac{\epsilon}{1-\epsilon}}x^{a_1\epsilon}\alpha_R$
and
$\beta\to\chi^{\frac{\epsilon}{1-\epsilon}}x^{b_1\epsilon}\beta_R$.} 
\be
U=\frac{\lambda_R}{4}\left[h^{\frac{2-\epsilon}{1-\epsilon}}
x^{a_1
\epsilon}-\zeta_R^2\chi^{\frac{2-\epsilon}{1-\epsilon}}
x^{b_1 \epsilon}\right] ^2\;~,
\label{ndim1}
\ee
and introduce counter-terms for all
terms appearing in the potential:
\ba
\label{cc1}
U_{cc}\!=\!
\Big[\!\!\!&&A\left(\frac{1}{\bar\epsilon}+a\right)h^{\frac{
2-\epsilon } { 1-\epsilon}}
\chi^{\frac{2-\epsilon}{1-\epsilon}}
x^{(a_1+b_1) \epsilon}+\\ 
&&B\left(\frac{1}{\bar\epsilon}+b\right)\chi^{\frac{4-2\epsilon}{
1-\epsilon}}
x^{2b_1 \epsilon} + 
C
\left(\frac{1}{\bar\epsilon}+c\right)h^{\frac{4-2\epsilon}{1-\epsilon}
}
x^{2a_1
\epsilon}\Big] ~\!\!.
\nonumber
\ea
As before, we do not introduce any modification of the kinetic terms.
Now we have more freedom in comparison with the GR-SI prescription 
due to the existence of new arbitrary parameters $a_1$ and $b_1$. 

The coefficients $A,~B,$ and $C$ are fixed as in Eq. (\ref{ABC}). The
parameters $a_1$ and $b_1$ can be chosen in such a way that the
one-loop effective potential does not contain terms $\chi^6/h^2$ and $h^6/\chi^2$, which are singular at $(0,0)$. These conditions lead
to  $a_1=0,~b_1=0$. Then the requirement that  the classical flat
direction $x_0=\zeta$ is not lifted by quantum effects gives 
(for $\zeta \lll 1$):
\ba
\nonumber
&&b=3a-7+2\log(2\lambda_R)+{\cal O}\left(\zeta_R^2\right)\\
&&c=\frac{1}{3}\left[a+7-2\log(2\lambda_R)
\right]
+{\cal O}\left(\zeta_R^2\right)~.
\ea

With all these conditions satisfied the  one-loop effective potential
as a function of $h$ for $\chi=\chi_0$ fixed, $h ~\sim~ \zeta
\chi_0 =v$ and $\zeta\lll 1$ is {\em different} from that in
Eq.~(\ref{pp}):
\be
U_1 = \frac{m^4(h)}{64\pi^2}\left[\log\frac{m^2(h)}{v^2}
+{\cal O}\left(\zeta_R^2\right)\right]
+ P_1 \log\frac{h^2}{v^2} + P_2~,
\label{pp1}
\ee
where  $P_1,~P_2$ are quadratic polynomials of $h^2$ and $v^2$.
Though the first term  is exactly the standard effective potential for
the theory (\ref{L}) with the dynamical field $\chi$ replaced by a
constant $\chi_0$, the rest is not simply a redefinition of the
coupling
constants of the theory due to the presence of $\log\frac{h^2}{v^2}$.
In other words, even the low energy physics is
modified in comparison with ordinary renormalizable theories.

\end{document}